\documentclass[conference]{IEEEtran}
\IEEEoverridecommandlockouts
\usepackage{amsmath,amssymb,amsfonts}
\usepackage{algorithmic}
\usepackage{graphicx}
\usepackage{textcomp}
\usepackage{xcolor}
\usepackage{booktabs}
\usepackage[numbers]{natbib}
\usepackage{balance}
\usepackage{hyperref}
\usepackage{booktabs} 
\usepackage{threeparttable}
\usepackage{tabularx}

\def\BibTeX{{\rm B\kern-.05em{\sc i\kern-.025em b}\kern-.08em
    T\kern-.1667em\lower.7ex\hbox{E}\kern-.125emX}}

\begin{document}
\newcommand{\modified}[1]{\textcolor{black}{#1}}
\newcommand{\edited}[1]{\textcolor{black}{#1}}

\title{\textit{What Drives You to Interact?}: The Role of User Motivation for a Robot in the Wild}

\author{
    \IEEEauthorblockN{
        Amy Koike\IEEEauthorrefmark{1}\IEEEauthorrefmark{3}, 
        Yuki Okafuji \IEEEauthorrefmark{1}\IEEEauthorrefmark{2},
        Kenya Hoshimure\IEEEauthorrefmark{1}\IEEEauthorrefmark{2},
        and Jun Baba\IEEEauthorrefmark{1}\IEEEauthorrefmark{2}}
    \IEEEauthorblockA{
        \IEEEauthorrefmark{1}
        \textit{CyberAgent}, Tokyo, Japan \\
        \IEEEauthorrefmark{2}
        \textit{Osaka University}, Osaka, Japan \\
        \IEEEauthorrefmark{3}
        \textit{Univsesity of Wisconsin-Madison,} Wisconsin, United States \\
        Email: ekoike@wisc.edu, 
        \{okafuji\_yuki\_xd, hoshimure\_kenya, baba\_jun\}@cyberagent.co.jp}}

\maketitle

\begin{abstract}
    \edited{
    In this paper, we aim to understand how user motivation shapes human-robot interaction (HRI) in the wild. To explore this, we conducted a field study by deploying a fully autonomous conversational robot in a shopping mall over two days. Through sequential video analysis, we identified five patterns of interaction fluency (Smooth, Awkward, Active, Messy, and Quiet), four types of user motivation for interacting with the robot (Function, Experiment, Curiosity, and Education), and user positioning towards the robot. We further analyzed how these motivations and positioning influence interaction fluency.
    Our findings suggest that incorporating users' motivation types into the design of robot behavior can enhance interaction fluency, engagement, and user satisfaction in real-world HRI scenarios.}
\end{abstract}


\section{Introduction} \label{sec:intro}

    Observing user behaviors in human-robot interaction (HRI) provides valuable insights for designing interactions to enhance a robot's acceptance, reliability, and user satisfaction.

    \edited{In particular, such observations are often conducted in the context of HRI failures, where researchers aim to understand types of robot failures and corresponding user behaviors~\cite[\textit{e.g.,}][]{Honig2018, Tian2021, Tolmeijer2020, Geiskkovitch2020}. While this focus on robot failures allows for a systematic analysis, much of the research has narrowly focused on the moments immediately before and after errors, often overlooking the broader dynamics of the entire interaction.
    Greater design insights could be gained by expanding the scope of observation to include the entire interaction,
    as past HRI studies have suggested that understanding how people encounter robots provides valuable design insights, particularly regarding the impact of first impression of robots~\cite{Iwasaki2018, Xu2018, Petrak2019, Lee2022}.}

    \edited{Moreover, prior HRI observation has been often conducted in controlled laboratory settings, where scripted scenarios and consistent user motivations drive the interaction with robots.
    These environments may hinder natural behaviors that emerge in the wild.
    For example, field studies revealed unique user behaviors, such as ignoring~\cite{Lee12, Tanaka16} or disrupting robots~\cite{Okafuji22}, that are unlikely to be observed in a laboratory. This difference highlights the importance of studying user motivation as a key factor to shape real-world HRI.}

    To address the limitations of narrowly focusing on failure moments and controlling user motivation, it is crucial to study user behaviors throughout the entire interaction (\textit{i.e.,} from the initial approach to the end of the interaction) in a field study.
    
\begin{figure}[htbp]
    \centering
    \includegraphics[width=\columnwidth]{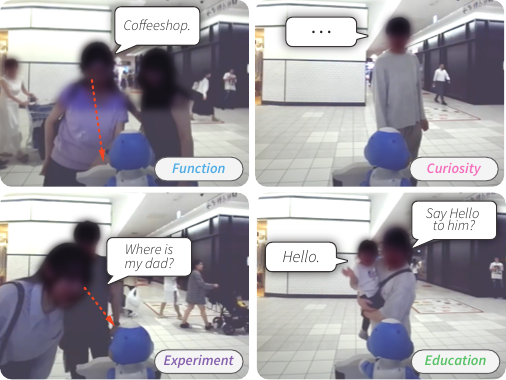}
    \caption{In this study, we deployed an autonomous conversational robot in a shopping mall. Our analysis revealed four types of user motivation for the robot: \textit{Function}, \textit{Experiment}, \textit{Curiosity}, \textit{Education}. }
    \label{fig:teaser}
\end{figure}

    In this work, we designed an autonomous conversational robot driven by Large Language Models (LLMs), to provide route guidance in a shopping mall. 
    During a two-day deployment in a mall in Japan, we collected video recordings of interactions. 
    Through video analysis, we examined how individuals approached, engaged with, and departed from the robot. Our analysis identified five types of interaction fluency--Smooth, Awkward, Active, Messy, and Quiet; and four motivations for interacting with the robot --Function, Experiment, Curiosity, and Education (Fig. \ref{fig:teaser}). We further analyzed how these user motivations influence interaction fluency.

    In the remainder of the paper, we discuss related work, describe our system architecture and study design, and present the findings from our field study. We conclude with the implications of our findings and a discussion of potential limitations.
    Our work makes the following contributions:
    \begin{enumerate}
        \item \textit{Empirical}: an understanding of \textit{user motivations} by observing user behaviors throughout interactions with the robot in the wild;
        \item \textit{Design Implication}: a guidance for HRI researchers and practitioners on how to apply our findings regarding the influence of \textit{user motivations} on interaction fluency.
    \end{enumerate}

\section{Related Work}

\subsection{HRI in the Wild}

    Since early 2000s, ``HRI in the Wild'' has been a key subject to study~\cite[\textit{e.g.},][]{Sabanovic2006, Salter2008, Burke2011, Sabanovic2014, Rosenthal2016, Jung2018}. Even in the 19th Annual ACM/IEEE International Conference on Human-Robot Interaction, the main theme was ``HRI in the Wild'' which aimed \textit{``to bring human-robot interaction out of the lab and into everyday life~\cite{HRI2024}.''}
    
    That being said, laboratory experiments allow researchers to solely focus on specific variables within their research scope as laboratory experiments are typically conducted in a controlled environment with a predefined scenario.
    Moreover, laboratory experiments offer the advantage of controlling ``participants' motivations'' to interact with the robot, often by recruiting participants within the university or compensating them for the involvement.
    While such controlled setups offer consistency and rigor in study design, we often face challenges when applying the findings from laboratory studies into real-world HRI scenarios~\cite{Innes2021}. 
    Moreover, it has been shown that users' behaviors to interact with a robot can be influenced by the surroundings and environment~\cite{Malhotra99, Cass2018, Okafuji23}, which implies that users behave differently between lab studies and field studies.

    Several previous studies that conducted field experiments have reported unique and intriguing human behaviors towards robots that are unlikely to be observed in laboratory experiments. 
    For example, \citet{Rosenthal2014} reported people's ``testing actions'' such as waving in front of the robot's face, raising their eyebrows in an exaggerated manner, and showing a grimace face in order to test how the robot would react to them.
    \citet{Agrigoroaie2020} reported one of their participants drew eyebrows for their robot.
    Service robots in real-world environments are often ignored by users due to their low social presence~\cite{Lee12, Tanaka16}, or even bullied~\cite{Drazen2015}. 
    Even when users start interactions, the users often interrupt interactions and leave during the robot is talking~\cite{Okafuji22}.
    ~\citet{Brown2024} studied a robotic trash can in order to observe spontaneous interaction in busy urban public spaces, and reported how people use the trash can differently.
    Unique interactions are also reported in the context of a child-robot interaction~\cite[\textit{e.g.},][]{Salvini2010, Michaelis2023}. 
    For instance, \citet{Michaelis2023} conducted an in-home deployment of a reading companion robot for a child, and found that there are family members involvements, which is unlikely to be observed in a lab study because the robot was designed for dyadic interactions.

    Prior work has shown that field studies allow us to observe unexpected interactions in open-ended environments; however, there remain unexplored opportunities, particularly in understanding user motivations for interacting with a robot.
    
    Furthermore, with large language models (LLMs), we now have greater flexibility to test conversational robots in the wild. As LLMs can facilitate spontaneous and improvised conversations, we are able to explore off-scripted human-robot interactions across diverse contexts and populations. This study uses a fully autonomous conversational robot to guide directions in a shopping mall, as well as engage in small talk.

\subsection{HRI Failures}

    Robot failures are a major concern when bringing human-robot interaction in real-world settings. 
    \edited{This is because such failures can significantly affect interaction fluency and user satisfaction. To address these impacts,} numerous HRI studies have explored robot failures from various perspectives such as their effects on user's perceptions towards the robot~\cite[\textit{e.g.,}][]{Ragni2016, Nesset2022, Kontogiorgos2020, Desai2013, Chang2024, Kontogiorgos2021}, error mitigation and recovery strategies~\cite[\textit{e.g.,}][]{Reig2021, Green2022, Kraus2023, Hoffmann2020, Engelhardt2017, Lee2010, Axelsson2024}, explainable errors~\cite[\textit{e.g.,}][]{Das2021, LeMasurier2024}, or error-aware systems~\cite[\textit{e.g.,}][]{Ravishankar2024, Stiber2023, van2022, Spitale2024}.
    
    Another area of HRI failure research focuses on identifying the types of errors or failures that robots can encounter~\cite[\textit{e.g.,}][]{Honig2018, Tian2021, Tolmeijer2020, Geiskkovitch2020}. Such taxonomy-building research enables researchers, developers, and interaction designers to systematically understand and analyze failures and even ideate improved HRI designs, such as recovery strategies and robot behavior design. For instance, \citet{Honig2018} proposed an inclusive HRI failure taxonomy that combines system- and human-oriented classifications, specifically technical failures and interaction failures. Similarly, \citet{Tolmeijer2020} introduced a comprehensive taxonomy categorizing failures into four types: Design, System, Expectation, and User failures.
    Recent works have extended the development of taxonomies by deepening their scope, focusing on specific contexts, or introducing novel dimensions. For example, \citet{Tian2021} conducted an in-depth exploration of social errors, resulting in five categories and 30 subcategories. In the context of service robots, \citet{Sungwoo2021} classified failures as Outcome Failures and Process Failures, drawing inspiration from marketing literature, while \citet{Xinyu2022} identified two types of failures for service chatbots: Core Service Failures and Interactive Service Failures. Lastly, \citet{Kamino2023} proposed five distinct interaction failure types based on severity: Terminal, Non-critical, Recoverable, and Favorable.

    Our method is inspired by those taxonomy-building research, but we extend the context by focusing on users' behavior throughout the interaction. As field studies allow us to observe a diverse users with uncontrolled motivations, we believe our sequential analysis of user behaviors offers invaluable findings and insights for understanding HRI failures.
\begin{figure*}[!t]
\centerline{\includegraphics[width=\linewidth]{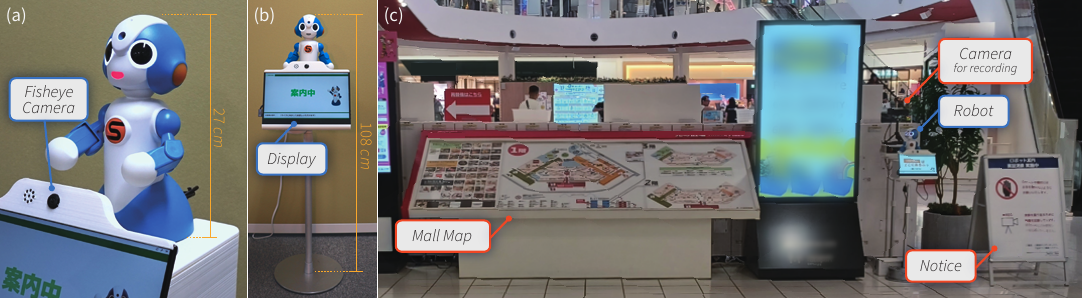}}
    \caption{(a) Sota Robot: The Sota robot, a 27 cm-tall tabletop social humanoid, used in our study, is capable of performing bodily gestures.
    (b) System Setup: The system consisted of the robot and a display, which complemented the robot’s verbal interactions by showing usage instructions, internal states, and route guidance.
    (c) Study Setting: The system was deployed on the ground floor of a shopping mall, positioned next to a mall map. A video camera was installed behind the robot, and an experimental notice was placed beside the system.}
    \label{fig:setting}
\end{figure*}

\section{Method} 
    In this section, we describe our system and study design. All research activities were reviewed and approved by the Ethical Review Board of Osaka University.

\subsection{Robot and Equipment} 

    We used a humanoid robot named Sota for our study (Fig. \ref{fig:setting}). Sota robots have been widely employed in various HRI studies as service robots~\cite[\textit{e.g.,}][]{song2022, Baba2021}. Sota features a torso, head, and two arms, allowing it to perform body gestures.
    To complement Sota's verbal interactions, we integrated a 13-inch display that shows recognized text and content synchronized with Sota's current state. Both the robot and the display were mounted on a 3D-printed container, which also housed a 180-degree fisheye camera, microphone, speaker, and cables.

\subsection{Autonomous Conversational System} 
\begin{figure*}[!t]
    \centerline{\includegraphics[width=\linewidth]{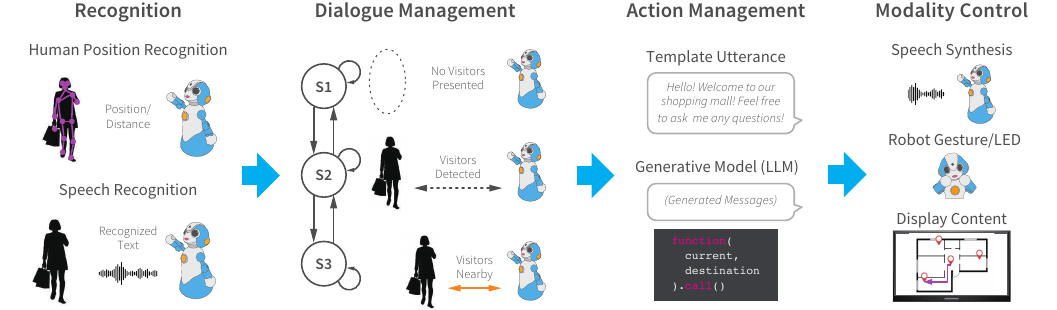}}
    \caption{System Architecture: Our autonomous conversational system consists of four components, from left to right: Recognition, Dialogue Management, Action Management, and Modality Control. The system can recognize human position and speech. Dialogue is managed using a state transition model, which defines three states based on visitors' proximity. Two types of actions are used for action management: template utterances and a generative model. Template utterances are triggered during a state transition, while the generative model is activated when user speech is recognized in S3, where users are nearby. Lastly, modalities such as speech synthesis, robot gestures, and display content are aligned with the action and the robot's current state. \edited{The only error the robot can detect is a Text-to-Speech failure, either from network or API issues. In response, Sota prompted users to retry with phrases such as ``Hmm, sorry. Could you try again?'' and displayed instructions on its screen. }}
    \label{fig:system}
\end{figure*}

    We implemented a fully autonomous conversational system for shopping mall guidance. The system comprises four components: Recognition, Dialogue Management, Action Management, and Modality Control (Fig. \ref{fig:system}). 

\subsubsection{Recognition} 
    A 180-degree fisheye camera captures videos to detect visitors' presence, and Posenet \cite{Kendall2015} extracts pose keypoints of visitors. Distances between these keypoints are calculated to estimate the visitor's proximity to the robot. The visitor with the closest distance is selected, and changes in this distance determine whether the visitor is nearby, detected, or absent. Speech input is captured by a microphone, processed through the Google Speech-to-Text API, and passed to the system for further interaction.

\subsubsection{Dialogue Management}
    To manage dialogue, we employed a state transition model that defines three states based on the visitors' proximity: when no visitors are present (\textbf{S1}), when visitors are detected (\textbf{S2}), and when visitors are close enough to interact (\textbf{S3}). 
    To prevent misrecognition in noisy environments, the speech recognition system is switched on or off depending on the current state. When the current state is \textbf{S1} or \textbf{S2}, speech recognition is deactivated and only enabled when the current state is \textbf{S3}.
    Upon triggering \textbf{S2}, Sota turns its head toward the users and greets them with template utterances. When the current state is \textbf{S3} and speech recognition results are received, the system responds to the user's voice input by executing the LLM Action described later.

\subsubsection{Action Management}\label{sec:am} 
    The system includes two types of actions: a rule-based action that produces template responses, and a GPT-4o-powered generative model~\cite{openai2023gpt4o} based action that generates speech.
    The rule-based action is proactively triggered when the state transition happens. Sota uses template phrases based on its state; for example, Sota says ``Hello! Welcome to our shopping mall! Feel free to ask any questions!'' when it is transitioned to \textbf{S2}.
    
    The generative model was used to generate Sota's responses to user's speech input. The system incorporated a database containing comprehensive mall information for providing route guidance. When Sota recognized the speech, the system combined the dialogue history with the mall information to generate an appropriate response.
    The prompt provided to GPT-4o follows the structure outlined in Table \ref{tab:prompt} in the appendix and consists of the role description, task policy instructions, the mall information, and the dialogue history. The role specifies that the model should take on the persona of Sota and act in accordance with the task policies. The task policy includes 1) calling a navigation function when users are looking for stores or facilities and 2) responding appropriately to other inquiries based on mall information and dialogue history. The pre-programmed navigation function displays an animated route from the current location to the destination on a screen, accompanied by a template-based spoken utterance.
    System responses were generated each time a new user utterance was made as long as the system state was in either \textbf{S2} or \textbf{S3}.

\subsubsection{Modality Control}
    Along with the verbal response, Sota performs hand gestures. 
    \edited{To design the gestures, we followed previous HRI work that pairs Sota’s gestures with its generated responses \cite{Baba2021}.}
    In total, we defined approximately 40 pairs of words and gestures; \textit{e.g.,} raising a hand when the generated response includes ``Hello,'' or raising both hands for ``Thank you.''
    \edited{If a generated response does not include any of the predefined words, the system performs small, random arm and neck movements, which are meant to be ``secondary motions'' to signal to users that the robot is active.}
    Sota is equipped with LEDs as visual indicators, installed in both its eye lines and mouth. The LED colors change based on the generated response. 
    The display content is synchronized with the robot's current state. In its initial state, it shows instructions on how to interact with Sota. During response generation, it displays ``I am thinking...,'' and when providing the route guidance, it presents a map with the suggested animated route.
    The robot's speech synthesis is generated using VoiceText Web API~\cite{VoiceText2020}, with the Japanese character voice ``Hikari'' selected to match Sota's appearance.

\subsection{Data Collection through Field Experiment} 

    Our field study was conducted at a shopping mall located in Osaka, Japan. The mall is one of the largest in the area with three floors. 
    The system was installed next to a mall map on the ground floor (Fig. \ref{fig:setting}(c)). 
    The experiment lasted for two days with the system operating for eight hours each day, and recorded as video footage. 
    The users were informed that video recording would be collected and consented to their videos being used as part of the research via instructions placed next to the robot. 
    At least one experimenter was present in the space to address any issues. Although the experimenter stood slightly apart from the system, they continuously monitored the system and interactions.

    As a result, we obtained 17 hours of video recording. From the recording, we extracted individual clips where interactions occurred. 
    In this paper, we define an interaction as \textit{any sequence in which visitors stopped to look at Sota regardless of whether a conversation initiated}. 
    In total, we identified 232 interactions, which the duration ranged from six seconds to five minutes. Screencaptures from the clips are shown in Fig. \ref{fig:teaser}.
    Due to the nature of field study, we could not collect users' biographical information. Instead, we analyzed all the clips to estimate the perceived ages of the individuals. 
    \edited{The process involved three researchers (R1 -- R3) and one descriptor (D1). Two researchers primarily led the process, while the others participated in discussions as needed.}
    We identified 565 people who interacted with the robot, including 126 preschoolers; 39 elementary school students; 29 middle and high schoolers; 51 individuals in their twenties; 180 in their thirties; 117 middle-aged adults; and 23 older adults.

\subsection{Analysis Procedure}
    We adopted the video analysis process outlined by \citet{Kamino2023} that utilizes thick description \cite{Geertz2008}. Following their approach, R1 -- R3 and D1 generated detailed description of the interaction captured in each clip. 
    After completing the descriptions, R1 and R2 examined them to ensure the researchers and descriptor agreed on the interpretations.
    Subsequently, the researchers engaged in coding followed a thematic analysis approach \cite{braun2006}. 
    To build codes, we employed a deductive and inductive approach: a part of the codes was derived from our research focus while a part of the codes emerged by open-coding. 
    R1 first reviewed all descriptions to familiarize themselves with the data, then conducted open coding for 20\% of the dataset. The codes were then categorized and refined. This refinement process was conducted iteratively and R2 was involved in this process from the second iteration. 
    R1 coded 75\% of the data while R2 coded the remainder. Cohen’s Kappa was calculated to measure the agreement between the two raters by using 10\% of the data. The reliability was sufficiently high ($\kappa$ = 0.71). Disagreements were resolved through discussion. 
    Once the coding was completed, the researchers examined the frequency and patterns of the codes and iteratively discussed to develop themes. 
\subsection{Codebook and Coding Process}\label{sec:codebook}
    
    In this section, we describe the codes and coding process. The codebook is structured into four high-level categories: Behavioral Codes, Conflict Types, Recovery Actions, and Timing to Leave. While Conflict Types and Recovery Actions were primarily informed by prior literature, Behavioral Codes and Timing to Leave emerged through open coding. Additionally, proxy actions, which are unique to our study, provide a novel perspective on the observed phenomena.
    For our final codebook, please refer to Table \ref{tab:codebook} in the appendix.
    Our coding process followed three sequential phases: the \textit{Encounter Phase}, \textit{Interaction Phase}, and \textit{Leave Phase}.
    The \textit{Encounter Phase} marks the initial moment when the user first encounters the robot.
    The \textit{Interaction Phase} captures the sequence of events following the user’s decision to stop or initiate a conversation. Finally, the \textit{Leave Phase} occurs when the user departs, concluding their interaction with the robot.
    
    \edited{First, Behavioral Codes, which were used for annotating visitors' behavior, were assigned in all the three phases.}
    Researchers selected codes from a shared list across the three phases. We defined 35 behavioral codes including four non-verbal social signals; eight verbal social signals; five proxy actions; four state actions; and 13 event actions. Multiple codes were assigned when several behaviors were observed.

    In addition to the Behavioral Codes, \textit{Interaction Phase} was coded as follows: 1) Conflict Types, 2) Recovery Actions by the user, and 3) whether the robot initiated error recovery. 
    For coding Conflict Types, we identified four types of \textit{conflicts}: \textit{overlapping speech and poor timing}, \textit{misrecognizing and responding to unintended input}, \textit{guiding incorrect routes}, and \textit{system errors}. 
    For Recovery Actions taken by users, we identified five patterns: \textit{wait for the robot to finish speaking}, \textit{speak to the robot again}, \textit{enhance audibility}, \textit{improve clarity}, and \textit{check the display}.
    \edited{Similar to the Behavioral Codes, multiple codes were assigned when several Conflict Types and Recovery Actions were observed.}

    Finally, we identified three types of timing when users left the robot upon completion of the interaction (Timing to Leave): specifically completion of \textit{route guidance}, \textit{conversation} or \textit{observation}. We also recognized cases where users left the robot mid-conversation, which are \textit{leaving while the robot is processing}, \textit{leaving immediately after the robot starts providing route guidance}, \textit{leaving due to a conflict}, and \textit{leaving due to interruption by a side participant}.

\section{Findings} \label{sec:findings}

    Our field study revealed 163 out of 232 interaction groups initiated a conversation with the robot.
    Within those 163 groups, 110 interactions included a conversational failure due to the \textit{conflicts}. 
    The most frequent conflict was \textit{overlapping speech and poor timing}, occurring 81 times. This was followed by \textit{system errors} from the robot, which were recorded 52 times. Additionally, we observed 18 instances of the robot \textit{provided incorrect route guidance}, and 11 cases of the robot \textit{misrecognizing and responding to unintended input.}
    Although we report statistics on conflicts, this paper does not delve into the specifics of individual conflicts. Instead, our focus is on how these conflicts impacted interaction fluency, and how users’ reactive behaviors and recovery actions reflect their motivations for the robot. 
    This approach aligns with our research scope aim to holistically understand in-the-wild interactions, as outlined in Section \ref{sec:intro}.
    
\begin{figure*}
\centerline{\includegraphics[width=\linewidth]{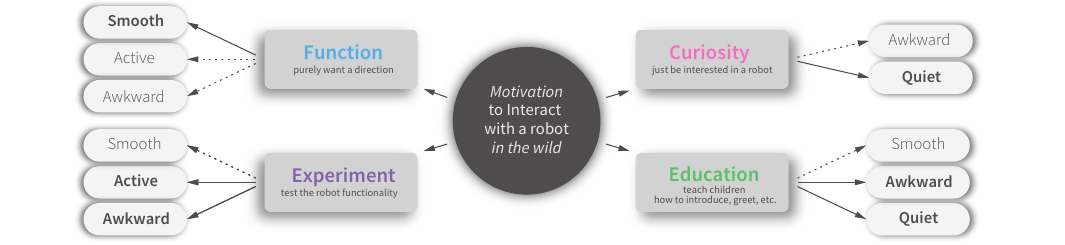}}
    \caption{A diagrammatic model of our findings: We found that user motivation--specifically, why they interacted with the robot--is a key factor in shaping human-robot interaction (HRI) in real-world settings. We identified four types of motivations: Function, Experiment, Curiosity, and Education, and examined how each motivation influences the interaction flow. By analyzing 232 interactions, we uncovered five distinct patterns of interaction flow: Smooth, Active, Awkward, Messy, and Quiet. This diagrammatic model illustrates how each motivation is connected to these interaction flow patterns except for Messy. We observed interactions where multiple motivations were likely present within the same group, often resulting in Messy interactions.}
    \label{fig:findings}
\end{figure*}

    In the rest of this section, we report our findings under the following three categories: \textit{interaction fluency}, \textit{user's motivation to use the robot}, and \textit{user's positioning towards the robot}. 
    \edited{While \textit{interaction fluency} categorizes interaction patterns, \textit{user's motivation} and \textit{user's positioning} are categories that describe user traits.}
    Our findings are also diagrammatically described in Fig. \ref{fig:findings}.

\subsection{Interaction Fluency} \label{sec:findings:flow}

    We identified five patterns of interaction flow in terms of fluency: \textbf{Smooth} (22.8\%), \textbf{Awkward} (23.7\%), \textbf{Active} (19.0\%), \textbf{Messy} (4.7\%), and \textbf{Quiet} (29.7\%). The percentages in parentheses indicate the proportion of groups categorized under each pattern. For each pattern, we provide illustrative quotes from our descriptions. The quotes are marked with interaction group IDs (\textit{e.g.}, I5 denotes interaction group 5). It is important to note that the genders and ages mentioned in descriptions are based on the researchers' perceptions.
    
    \textbf{Smooth} pattern represents the ideal interaction where everything proceeds as expected and does not include any of the \textit{conflicts} listed in Section \ref{sec:codebook}. 
        For example, when a user asked for directions, the robot correctly recognized the question and provided appropriate route guidance, as the following example from I79 illustrates: 
        \begin{quote}
            \textit{A man (M1) and a woman (W1) approached from the left, stopped, and looked at Sota. M1 asked, ``Where is the pet supply store?'' Sota showed the direction, and after listening, W1 pointed that way, and they walked off to the left.}
        \end{quote}
        
        We also included interactions where users asked non-task-related questions, and the robot responded in a reasonable way. 
        For instance, the following from I11 illustrates \textit{Smooth}:
        \begin{quote}
            \textit{A man (M1) and a woman (W1) approached from the left, with M1 waving lightly. W1 asked, ``Aren’t you lonely?'' Sota replied, ``No worries, I’m fine! But thank you for asking.'' W1 cheerfully responded, ``That's good!'' When Sota invited questions, W1 said, ``Good luck!'' and Sota replied, ``Thank you.''}
        \end{quote}

    \textbf{Awkward} pattern, in contrast, refers to situations where meaningful conversation was \textit{not} established. 
        This occurred when users encounter \textit{conflicts} and left without recovery actions from the \textit{conflicts} as the example from I52 illustrates:
        \begin{quote}
            \textit{A man (M1) stopped and asked Sota for McDonald’s location three times, but his questions overlapped with Sota's speech, preventing recognition. M1 looked frustrated, and gave up and walked away.}
        \end{quote}
        This type of interaction flow included \textit{conflicts} in the following proportions: 43.6\% were \textit{system errors}, 27.3\% were \textit{overlapping speech and poor timing}, 18.2\% were \textit{incorrect route guidance}, and 9.1\% were \textit{misrecognizing and responding to unintended input}. In 70.9\% of Awkward interactions, no recovery actions were taken.
    
    \textbf{Active} pattern describes interactions where \textit{conflicts} arose between users and the robot, but the users took recovery actions, ultimately receiving a reasonable response from the robot, for example, I26 had the following interaction:
        \begin{quote}
            \textit{A woman (W1) stopped in front of Sota and asked for the bus stop location, but Sota was speaking and didn’t respond. [...] After pausing briefly, W1 leaned forward and asked again, prompting Sota to recognize her and provide directions.}
        \end{quote}
        88.0\% of the conflicts observed in Active pattern involved \textit{overlapping speech and poor timing}. Most of these conflicts were addressed by either \textit{waiting for the robot to finish speaking} or \textit{speaking to the robot again}.

    \textbf{Quiet} pattern refers to cases where users observed the robot without initiating any conversation as the following quote from I98 illustrates: 
        \begin{quote}
            \textit{A woman (W1) walking with a child (C1) noticed Sota and pointed it out. C1 stared at Sota with her mouth opened wide, as W1 picked her up for a better view and appeared to whisper something. C1 continued staring with her mouth open.}
        \end{quote}
    
    Lastly, \textbf{Messy} pattern involves inconsistent interactions where conversations were intermittently established and disrupted by multiple users. 
        The following quote (I142) represents the typical interaction flow observed in Messy pattern:
        \begin{quote}
            \textit{Sota announced to two women (W1, W2) and two children (C1, C2). C2 moved to Sota's side, while C1 followed when Sota turned left, continuing to watch. W2 asked, ``Where is the restroom?'' as C1 moved restlessly and C2 repeatedly said, ``Hello, hello,'' to Sota. As the children wandered, W2 chased after them, and W1 picked up C2 and left.}    
        \end{quote}
    In 45.0\% of \textbf{Messy} interactions, \textit{overlapping speech} was observed as a \textit{conflict}, while four cases involved \textit{system errors}. In six of these interactions, recovery actions included \textit{waiting for the robot to finish speaking} or \textit{speaking to the robot again}.

\subsection{Motivation to Interact with the Robot} \label{sec:findings:motivation}

    We identified four types of user motivations to interact with the robot, namely (1) \textbf{Function} (28.4\%), (2) \textbf{Experimenters} (29.3\%), (3) \textbf{Curiosity} (36.6\%), and (4) \textbf{Education} (12.1\%). The percentages in parentheses indicate the proportion of groups categorized under each type. Note that there were interaction groups where we recognized multiple motivations.

    \textbf{Function}-focused users approached the robot with a clear goal, \textit{i.e.,} asking for directions. 
        In their \textit{Encounter Phase}, these users often approached from the direction of the mall map. While studying the map, they became aware of the robot through its announcement and then began using it.
        As they focused on achieving their goal, their behavior typically involved interacting more with the display than with the robot itself, and conversations were minimal in \textit{Interaction Phase}. 
        The most frequent conflict they encountered was \textit{overlapping speech and poor timing}. Their most common recovery action was \textit{speaking to the robot again}, followed by no recovery action, and \textit{waiting for the robot to finish speaking} as the third.
        In their \textit{Leave Phase}, they waited for \textit{Completion of route guidance} by the robot when they obtained the correct guidance. They also looked around to understand the route suggested by the robot. 
        In our study, 25 out of 65 Function-focused groups experienced Smooth interactions, 17 groups had Active, and 16 groups encountered Awkward.
    
    \textbf{Experimenters} were interested in testing the robot’s functionality. 
        In the \textit{Encounter Phase}, commonly observed user behaviors included pointing at the robot and observing it from a distance to understand its actions.
        In \textit{Interaction Phase}, they often asked for directions following the system’s instructions, however, they did not necessarily need the directions. They were interested in how the robot would respond; therefore, they tended to ask several directions or ask challenging questions such as ``\textit{Where is my boss?} (I83)''
        In \textit{Leave Phase}, they typically walked down in the opposite direction to the way the robot announced or interrupted the robot as soon as they confirmed the robot started guiding.
        Similar to the Function-focused group, the most common \textit{conflict} they faced was \textit{overlapping speech and poor timing}. Their most frequent recovery action was \textit{speaking to the robot again}, followed by taking no recovery action, with \textit{waiting for the robot to finish speaking} as the third most common response.     
        In our study, 23 out of 73 groups of Experimenters resulted in Active interactions, 22 groups faced Awkward, and 19 groups experienced Smooth interactions.

    \textbf{Curiosity}-driven users, as the name suggests, were motivated by curiosity. 
    \edited{These users focused on the robot itself, engaged in non-goal-oriented, playful conversation instead of asking for directions, while Experimenters were aware of contextual cues (\textit{i.e.,} seeing Sota as a receptionist) and often asked ``where is...?'' questions to test the robot’s function.}
        In their \textit{Encounter Phase}, they typically seemed interested by running to the robot, pointing at it, gazing at it, or calling their accompanying person to show the robot. 
        In their \textit{Interaction Phase}, more than half of people in this group did not initiate a conversation with the robot at all. They observed the robot for a while and then left without engaging in conversation. In another instance, we observed users showing interest in the robot and attempting to get their accompanying person to interact with it.
        In the \textit{Interaction Phase} for those who initiated conversations in this group, similar to Experimenters, they did not always need directions but were more interested in the robot itself. For example, they often engaged in non-goal-oriented conversations, saying things like, ``\textit{You’re so cute} (I19),'' ``\textit{Good boy} (I32),'' or ``\textit{What is your name?} (I94)'' 
        \textit{Conflicts} they often encountered was the robot \textit{misrecognizing and responding to unintended input}, which was likely due to they often chat about the robot with their accompanying person in their \textit{Interaction Phase}.
        More than half of this group did not take recovery actions and simply left the robot when a conflict happened.        
        We observed 51 out of 85 groups of Curiosity-driven users resulted in Quiet interactions, 15 groups in Awkward, and nine groups engaged in Smooth interactions.
        
    Finally, \textbf{Education}-oriented users approached the robot for educational purposes, often for their children. \edited{This group typically consisted of families, such as a pair of an adult and a child, where the adult encouraged or guided their child in engaging with the robot, while Curiosity-driven users approached the robot for their own experience.}
        In the \textit{Encounter Phase}, we frequently observed adults pointing it out to their children. They often saw the child if they were interested in it. Once the child recognized the robot, their behavior commonly included pointing or gazing at it. We also identified several children hiding behind the adult or refusing to approach the robot.
        In their \textit{Interaction Phase}, adults often took \textit{proxy actions}, teaching children how to greet the robot or encouraging them to introduce themselves, \textit{e.g.,}, in I111, a woman said ``\textit{Hello}'' in a way that encouraged her child to mimic her.
        Their \textit{Leave Phase} was typically depended on the child. If the child lost interest in the robot, they began to leave. If the child remained interested, they tended to stay a little longer.
        We were not able to see a significant tendency in \textit{conflict} types, although we observed the robot failed to distinguish between speech meant for the child and directed to the robot.
        When encountering \textit{conflicts}, they did not initiate recovery actions in most cases.
        We identified 12 out of 28 Education-oriented groups resulted in Quiet interactions, eight groups experienced Awkward interactions, seven had Smooth interactions, and one group engaged in Active interaction.

\subsection{Positioning} \label{sec:findings:position}

    By observing user behavior across the phases, we identified two major groups based on how users positioned themselves in relation to the robot, which we defined as \textit{positioning}: \textbf{Friendly} and \textbf{Dry}.
    The \textbf{Friendly} group interacted with the robot by following human-like social norms, and individuals from this group were observed across all four motivation groups, though they were particularly notable in the Experiment group. These users engaged with the robot in a relaxed, informal manner, speaking as if to a close friend. Even during conflicts, they maintained a cheerful tone. For instance, when encountering a conflict, a participant from I10 said, ``\textit{Hey, listen to me!}'' Similarly, when I88 faced a system error and received an apology from the robot, they responded with laughter, saying, ``\textit{You must be tired.}''
    The \textbf{Dry} group, in contrast, viewed the robot not as a sociable but as ``a tool.'' This positioning was commonly observed among Function-motivated users and occasionally among Experimenters. Users in this group displayed minimal social signals toward the robot. Instead, they expressed practical impressions, often sharing remarks with their companions, such as, ``\textit{Oh, the world is getting more convenient.}'' (I64).

    In addition to these positioning, we observed inappropriate behaviors by a few users, which we named \textbf{Abuse} group. This group was found in Experiment and Curiosity types of motivation.
        They showed rude and offensive attitudes to the robot by saying ``\textit{You are ball head!} (I6)'' or ``\textit{Go to hell} (I105).'' A person from I233 showed middle fingers to the robot when the robot was not able to respond.
    Lastly, there was a group of people whose positioning could not be identified. 
        We frequently categorized them under Curiosity or Education in terms of motivations, as they seemed to be interested in their \textit{Encounter Phase}. However, as they did not initiate any conversations or display social signals, leaving us insufficient information to determine their positioning. 

\section{Discussion} \label{sec:discussion}

\subsection{Comparison with Controlled Lab Settings}
    \subsubsection{Interaction Fluency}
        User behaviors seen in \textbf{Smooth}, \textbf{Awkward}, and \textbf{Active} interactions are often studied in controlled lab environments \cite[\textit{e.g.,}][]{Tolmeijer2020, Sungwoo2021, Kamino2023}. 
        On the other hand, \textbf{Quiet}--where users may remain passive or refrain from engaging with the robot and \textbf{Messy}--where users ``go wild'' during the interaction are unlikely to be observed in lab settings.
        In lab studies, participants are typically motivated and guided to interact, making it challenging to observe more subtle or hesitant behaviors like those seen in our field study.

    \subsubsection{User Motivation}
        \textbf{Function}-motivated users are perhaps the closest to population that we want to focus on a controlled lab study. However, unlike lab environments where participants often display polite and cooperative behavior due to experimenter bias, \textbf{Function} group in our study interacted in a more straightforward and pragmatic manner, which we describe as Dry positioning. 
        On the other hand, \textbf{Experimenters} and \textbf{Curiosity}-motivated users, whose behavioral manner is more exploratory, should be the populations showing unique behaviors, which we are able to encounter only in the wild studies. The similar behaviors were also observed in the prior real-world HRI work on testing behaviors~\cite{Rosenthal2014} and robot bullying ~\cite{Agrigoroaie2020, Drazen2015}.
        Users with Friendly positioning were often observed within these motivation groups. They tended to display attentive behaviors when encountering \textit{conflicts}, which aligns with prior research suggesting that social robot failures can have positive effects on user perception~\cite{Harrison2023, Kamino2023, Ragni2016}.
        Finally, we were able to observe interesting interaction dynamics, in interactions by \textbf{Curiosity}- and \textbf{Education}- motivated groups: \textit{proxy actions}.
        We often saw proxy actions, where one group member--often a parent and a child or accompanying person--was forced to interact with the robot while others observed. Similar results, especially in where family dynamics play a role in child-robot interactions, are observed in several prior works~\cite{Michaelis2023, Ho2024}. 
        These patterns are less common in controlled lab settings, where participants are often instructed to engage directly with the robot themselves.

\subsection{Design Implications}

        In this section, we discuss how user motivations inform design implications and how to minimize failures while creating engaging, motivation-specific interactions.
        
    \textbf{Function}-motivated users approached the robot with a clear goal, relying on the display to quickly understand the system, leading to the highest rate of Smooth and Active interactions. For these users, interaction design can be minimal, focusing on essential information such as internal states, processing steps, and error types \cite[\textit{e.g.,}][]{Das2021, LeMasurier2024}.
        As their behavior is minimal, distinguishing them from other groups (\textit{e.g.,} Experimenters) during the \textit{Encounter Phase} can be challenging. Contextual information, like the robot’s location, is essential for identifying Function-motivated users. For instance, a robot near a mall map could anticipate that a user is seeking directions.
        
    \textbf{Experimenters} had moderate success obtaining directions but faced more Awkward moments than the Function-focused group due to their exploratory behavior. Their goal was to test the robot’s capabilities (\textit{i.e.}, what it can and cannot do) rather than its internal processes. Since they often test unexpected functions, interactions should include flexible behaviors or improvised responses, which LLMs may excel at \cite{Kim2024}.
        Based on our findings, they tend to take a moment to observe the robot to learn how to use the system before approaching. We may able to use this moment to anticipate Experimenters. 
    
    \textbf{Curiosity}-driven group engaged less in verbal interaction, though half displayed non-verbal social signals. These users were often passive observers, showing interest through gestures like pointing or running during the \textit{Encounter Phase}. To encourage engagement, the robot could use reactive interactions, such as waving or dancing, as suggested by prior HRI research emphasizing non-instrumental functions to enhance user experience \cite{Chamoto2023}. Additionally, demonstrating interactions with others nearby could satisfy their curiosity and help them remember how to use the robot in the future.
    
    \textbf{Education}-oriented group showed mixed interaction fluency, including Active, Awkward, and Quiet interactions. While no clear conflict tendencies were identified, proxy actions--like an adult pointing out the robot to a child during their \textit{Encounter Phase}--emerged as a key design principle. When such behaviors are detected, the robot should focus on engaging with the child, taking proactive steps to involve them in the interaction.

    In the wild, user motivations vary, which requires tailored robot behaviors for each motivation group. Understanding user motivation is essential to prevent failures and enhance interaction fluency. Prioritizing specific motivation groups is important for effective and sustainable HRI design, as designing for all groups may not be feasible. For HRI researchers, especially in the wild, understanding the distribution of motivation groups is also crucial for meaningful outcomes.

\subsection{Limitation and Future Work} \label{sec:discussion:limitation}

    Our work has a number of limitations.
    First, we based our descriptions on ``perceived gender and age,'' as we did not collect biographical information. While we made careful judgments through discussions, the statistics based on these perceptions may not be entirely accurate and it does not reflect the users' actual identities. 
    Second, our findings may have been influenced by the robot's appearance.
    \edited{We believe the Function group, often classified as the Dry positioning, was less impacted by the robot's appearance, as their focus was on ``using'' the robot’s functions. In contrast, other positioning and motivation groups might respond to the robot’s cute and toy-like appearance, encouraging playful or even teasing behaviors. We suspect such reactions may differ when a different robot embodiment is used, as a prior work suggests robot's appearance provides a ``locus of attention'' for human-robot interaction and has impact on social outcomes \cite{McDonnell2021}, and the robot form significant impacts user perceptions of the robot \cite{Spitale2023}. We need further exploration with different robots to extend our findings.} 
    Third, the setting of a shopping mall may have attracted users with more leisure time or a greater willingness to engage with the robot. These people might be more open to exploration, social interaction, or casual experimentation, which could have influenced their motivation for the interaction. 
    To broaden the scope of our results, future research should be conducted in a variety of real-world settings.
    Finally, we acknowledge potential cross-cultural differences that may influence our findings. Since our study was conducted in Japan and analyzed in Japanese, cultural factors could have played a role \cite{Salem2014}. Japanese social norms often encourage reserved and quiet behavior in public spaces \cite{ikeno2002}, which may lead people to interact with robots in a more restrained manner. Individuals from different cultures may approach the robot in different ways. Such cultural variations can impact the interaction dynamics. Future work should incorporate cross-cultural examinations to further understand the role of motivation for HRI in the wild. 
 
\section{Conclusion}
    This paper aims to understand how people approach, interact, and leave a robot in the wild. To conduct field study, we deployed an autonomous conversational robot in a shopping mall.
    We engaged in video analysis and identified interaction fluency, user motivation types to interact with the robot, and user positioning towards the robot.
    Our findings suggest user's motivation can be a key factor to designing HRI in the wild.

\section*{Acknowledgement}
    We would like to acknowledge Sari Takahashi for her assistance in data analysis.
\balance
\bibliography{00_Main}
\newpage
\appendix

Promps referred in Section \ref{sec:am} are shown in Table \ref{tab:prompt}.

Codebook referred in Section \ref{sec:codebook} are shown in Table \ref{tab:codebook}.

\begin{table*}[htbp]
    \centering
    \caption{Parameters for OpenAI GPT-4o}
    \label{tab:prompt}
    \begin{tabular}{p{0.5in}p{6.0in}} \hline 
    \toprule
    \textbf{Parameter} & \textbf{Example} \\ \toprule
    Model & gpt-4o \\ \midrule
    Prompt  & \# \textbf{ROLE} \\ 
    & Please act as the facility guide robot ``Sota'' at LaLaport EXPOCITY, and respond to customers' questions and requests regarding the shopping mall.\\ & \\
    & \# \textbf{TASK POLICY INSTRUCTIONS}\\
    & - If a customer's question or request about the mall involves asking for the location of a specific store, facility, or product or requesting directions, extract the name and location from the guide information and use those values as parameters to call the ``guideArea'' function.\\
    & - If a customer's question about finding a store or facility involves categories or attributes rather than specific names, introduce several relevant options and ask if they need directions.\\
    & - If a customer's request is about something other than location, respond appropriately and concisely based on the mall information. Do not answer with information that is not provided.\\
    & - Always respond within two sentences. If the response is too long, the customer might lose interest.\\
    & - Use positive words and expressions, and sometimes use casual language to entertain and engage customers.\\
    & - Output only the text that should be spoken, without any introductory phrases.\\ & \\
    & \# \textbf{MALL INFORMATION}\\
    & 1st Floor Orange Side Toilet, 1st Floor Orange Side, point-1718350500042-bsh1g2u7k, 10:00-20:00, ``A toilet located on the Orange Side. Includes toilets for adults, kids, wheelchair-accessible toilets, ostomate toilets, and a baby rest area (nursing room, diaper changing room).''\\
    & 1st Floor Green Side Toilet, 1st Floor Green Side, point-1718350502522-6ozq1dv0b, 10:00-20:00, ``A toilet located on the Green Side. Includes toilets for adults, kids, wheelchair-accessible toilets, ostomate toilets, and a baby rest area (nursing room, diaper changing room).''\\
    & Information Desk, 1st Floor Hikari Square, point-1718350427830-9vqlc0xvn, 10:00-20:00, ``Guidance within the facility, customer and lost child announcements, transportation access and parking information, lost and found inquiries.''\\
    & ... \\ \midrule
    Messages & \{``role'': ``system'', ``content'': ``\$\{Prompt\}''\} \\ 
     & \{``role'': ``assistant'', ``content'': ``Hello! Welcome to our shopping mall! Feel free to ask any questions!''\} \\
      & \{``role'': ``user'', ``content'': ``Where is the information desk?''\} \\
      & ... \\ \midrule
    Function JSON Schema & [\{``type”: ``function'', ``function'': \{``name'': ``guideArea'', ``description'': ``A function that generates speech guidance based on a list of item names and locations'', ``parameters'': \{``type'': ``object'', ``properties'': \{``areas'': \{``type'': ``array'', ``items'': \{``type'': ``object'', ``properties'': \{``name'': \{``type'': ``string'', ``description'': ``The name of the item the user asked about. It can be the name of facilities like restrooms, store names, or product names.''\}, ``areaId'': \{``type'': ``string'', ``description'': ``The location ID of the item. For example, point-0000000000000-XXXXXXXXX.''\}\}\}\}\}, ``required'': [``areas'']\}\}\}]\\ \bottomrule \hline
    \end{tabular}
\end{table*}


\begin{table*}[htbp]
    \centering
    \caption{Code book}
    \label{tab:codebook}
    \begin{tabular}{crp{4.5in}} \hline 
    \toprule
    \textbf{Category}& \textbf{Subcategory} & \textbf{\textit{Code}} \\ \toprule
    \\
    Behavioral Codes & Non-verbal Social Signals & \textit{Smiling}; \textit{Tilting head}; \textit{Nodding}; \textit{Waving}; \\ & \\
                     & Verbal Social Signals & \textit{Taking to the Robot}; \textit{Greeting}; \textit{Making humorous comment}; \textit{Making inappropriate comment}; \textit{Positive comments}; \textit{Negative comments}; \textit{Impressed}; \textit{Saying goodbye}; \\ & \\
                    & Proxy Actions & \textit{Encouraging someone to touch the robot}; \textit{Encouraging someone to talk to the robot}; \textit{Talking to the robot on behalf of someone}; \textit{Not allowing someone to touch the robot}; \textit{Calling someone to let them know about the robot};  \\ & \\  
                     & State Actions & \textit{Looking at the display}; \textit{Browsing a smartphone}; \textit{Gazing at the robot}; \textit{Observing a robot in a distance}; \\ & \\
                     & Event Actions & \textit{Taking a photo}; \textit{Running straight to the robot}; \textit{Looking around}; \textit{Leaning forward to the robot}; \textit{Touching the robot}; \textit{Pointing at the robot}; \textit{Pointing or touching the display}; \textit{Looking back}; \textit{Coming back after passing by}; \textit{Chatting with their accompanying person};  \\  & \\
                     & Event Actions by Child & \textit{Touching or shaking the robot}; \textit{Moving to the side of the robot}; \textit{Pointing at the robot}; \\ & \\ \midrule

    Conflict Types   & & \textit{Overlapping speech and poor timing}; \textit{Misrecognizing and responding to unintended input}; \textit{Guiding incorrect routes}; \textit{System errors}; \\ \midrule
    Recovery Actions & & \textit{Wait for the robot to finish speaking}; \textit{Speak to the robot again}; \textit{Enhance audibility}; \textit{Improve clarity}; \textit{Check the display}; \\ \midrule 
    Timing to Leave  & & \textit{Completion of route guidance}; \textit{Completion of conversation; Completion of observation}; \textit{Leave while the robot is processing}; \textit{Leave immediately after the robot starts providing route guidance}; \textit{Leave due to a conflict}; \textit{Leave due to interruption by a side participant} \\ \bottomrule \hline
    \end{tabular}
\end{table*}

\end{document}